\newcommand{\beq}{\begin{equation}}
\newcommand{\eeq}{\end{equation}}

\documentclass{ws-procs9x6}
\begin{document}

\title{Hints on dual variables from the lattice SU(2) gluodynamics}

\author{V.~I. ZAKHAROV}

\address{Max-Planck Institut f\"ur Physik \\
Werner-Heisenberg Institut \\ 
F\"ohringer Ring 6, 80805, Munich\\ 
E-mail: xxz@mppmu.mpg.de}

\maketitle

\abstracts{In many cases, topological excitations become 
fundamental variables in a dual 
formulation of the theory.
Assuming this is true in case of $SU(2)$ gluodynamics, we look for
hints on the dual variables from the  lattice simulations.
Two-dimensional vortices,
or branes with recently established properties seem to be most
natural candidates. The total area of the branes scales in the
physical units while the non-Abelian action is ultraviolet divergent.
The branes are populated with magnetic monopoles, or tachyonic mode.}

\section{Introduction}

Duality proves a powerful tool to get insight into dynamics
of a system, for a review see, e.g., \cite{savit}. A recent landmark
is a proof of the confinement using duality within certain supersymmetric
Abelian field theories  \cite{seiberg}.
There exist also other examples of Abelian gauge theories 
which allow for dual formulations in terms of electric and
magnetic variables \cite{savit}. As for non-Abelian
theories, it has become clear
that search for dualities 
goes in this case beyond field theories,
 see, in particular, \cite{maldacena}. 
This makes the search for dualities even a more profound problem.  

Search for dual formulations is a  theoretical challenge, first of
all. Here we focus on a much less ambitious aspect of the problem, 
trying to understand what kind of hints on
the dual variables can be extracted from
the lattice simulations
of $SU(2)$ gluodynamics. The point is that 
topological excitations found within  the original
formulation can become dual variables, see, e.g., \cite{savit}.
Let us explain this on the Seiberg-Witten example. In $N=2$
SUSY Abelian theory there exist classical
monopole solutions with 
mass equal to
\beq\label{bps}
M_{mon}~=~\langle \phi\rangle \cdot Q_M~~,
\eeq
where $\phi$ is a scalar filed 
 and $Q_M$ is the magnetic charge which is inverse
proportional to the electric charge, $Q_M=const/e$. Moreover, the
solution saturates the BPS bound. The $N=2$ supersymmetry ensures
absence of quantum corrections to the mass. Thus, one can prove that
the monopole is becoming light for large original couplings $e$. This
observation can be considered as a hint that there exists a dual
formulation in terms of the monopoles.

In case of gluodynamics, effective degrees of freedom responsible for the
confinement are natural candidates for future dual variables. 
Phenomenologically, magnetic monopoles and P-vortices
are most promising, for a recent review see \cite{greensite}.
Note that both monopoles and P-vortices condense which is a common
sign of the dual variables.

In this paper we concentrate on field theoretical properties of the monopoles
and vortices. Trying to appreciate the anatomy of the monopoles
in terms of the Yang-Mills fields, we note
first that apriori it would be easy to argue
that the monopoles might not play any dynamical role. Indeed, there
are two known types of Abelian monopoles, corresponding to singular and 
non-singular fields at the origin. We argue that generically
both types of the Abelian monopoles 
cannot be generalized to the non-Abelian case. The only possibility
left is a self tuning of the suppression due to the action and
of enhancement due to the entropy. Both should be ultraviolet
divergent but cancel each other to the order $\Lambda_{QCD}$.

There is no known mechanism for the self-tuning in the original
Yang-Mills theory and the possibility looks pure
academic. Amusingly enough, lattice data do provide
strong evidence for such self-tuning.
Theoretically, one can then go further and argue that self-tuning of
the monopoles is not sufficient,
even if it is granted by the data, and monopoles should belong
to self-tuned two-dimensional surfaces. The data turn again to support this
conclusion and  
the surfaces in point are
the  P-vortices, for review see
\cite{greensite}. The P-vortices are indeed closely associated with
monopoles. 
What makes the self-tuning manifest is a recent observation
that the P-vortices and monopoles have non-Abelian action
which is divergent in the ultraviolet 
\cite{anatomy,kovalenko,syritsyn}.
In other words,  
the lattice spacing $a$ is seen in 
distribution of the non-Abelian action
on and around the vortices. On the other hand, $\Lambda_{QCD}$
controls the monopole mass and vortex string tension.

We will call these thin two-dimensional surfaces populated
with tachyonic mode (monopoles) branes. (The term vortex is
somehow reserved, in the studies of confinement,
 for thick, or bulky objects, see, e.g., 
\cite{greensite,tomboulis}.)   
In Sect. 2 we discuss self-tuning of the magnetic monopoles in
the lattice $SU(2)$. In Sect. 3 we describe evidence in favor
of self-tuning of the vortices. Conclusions are in Sect. 4.
Note that the argumentation outlined above was already 
presented partly in Refs. \cite{interpretation,vz}.

\section{Self-tuning of the lattice monopoles}

\subsection{Fine tuning of the Abelian monopoles}

The size of the BPS monopole (\ref{bps}) is large, that is of order
$\langle\phi\rangle^{-1}$. The softening of the fields  
is due to a non-trivial scalar field configuration. 
Now we turn to another well understood case
of the monopole condensation, that is compact $U(1)$,
for  review see, e.g., \cite{peskin}.
Generically, the idea is the same as above:
condensation of the monopoles is signaled by vanishing of the monopole
mass. There are two important changes, however. First, there is no scalar
field so that magnetic field is very strong at short distances.
Second, lattice theories are considered in the Euclidean space.

In more detail, we consider electromagnetic field on the lattice.
In the continuum limit the Lagrangian is that of free field:
\beq
L~=~{1\over 4e^2}F_{\mu\nu}F_{\mu\nu}~~.
\eeq
where $F_{\mu\nu}$ is the field strength tensor.
Since there is no softening Higgs field the radiative
mass
of the monopole is divergent  in the ultraviolet:
\beq\label{uv}
M_{rad}~=~{1\over 8\pi}\int_a^{\infty}d^3r|{\bf H}|^2~\approx~
{const\over e^2}{1\over a}~~,
\eeq
where $a$ is an ultraviolet cut off, ${\bf H}$ is the magnetic field,
$|{\bf H}|\sim 1/(er^2)$ and $e$ is the electric charge appearing
because  of the Dirac quantization condition.

Because of the ultraviolet divergence (\ref{uv}), 
the monopoles can be considered
point-like. The probability to observe a monopole trajectory of the
length $L$ is suppressed then by the action as
\beq\label{action}
exp(~-S)~\sim~\exp\big(~-{const\over e^2}{L\over a}\big)~~,
\eeq
and at first sight the monopoles are removed from the physical
spectrum in the limit $a\to 0$.
Remarkably, this conclusion  is actually not true. Indeed, the
probability $W(L)$ to observe a trajectory of length $L$ is a product of
(\ref{action}) and the entropy factor $N_L$
which is the number of ways the length $L$
can be realized on the lattice. The latter is also ultraviolet divergent
and:
\beq\label{suppression}
W(L)~\sim ~\exp(-M_{ren}\cdot L)~~,
\eeq
where the renormalized mass is approximately:
\beq
M_{ren}~\approx~{1\over a}\big({const\over e^2}-\ln 7\big)~~,
\eeq
where the $\ln 7$ term is of pure geometrical origin, due to the
entropy.

Monopole condensation occurs when any trajectory length is not suppressed by 
(\ref{suppression}). In other words, choosing $e^2=e^2_{crit}$, where
\beq\label{condition}
{const\over e_{crit}^2}-\ln 7~=~0~~,
\eeq
ensures condensation of the monopoles.
Validity of this condition was checked numerically,
see \cite{shiba} and references therein.

More generally, starting with the classical action for a point-like
particle:
\beq
S_{cl}~=~M(a)\cdot L~~,
\eeq
one can develop field theory in the Euclidean space-time
in the polymer representation,
see, e.g. \cite{sym}. In particular, evaluating the propagator 
as a path integral a la Feynman one arrives at the following
expression
for the propagating, or physical mass:
\beq\label{physical}
m^2_{phys}~\approx~{8\over a}\big(M(a)-\ln 7/a\big)~~,
\eeq
where the factor  $\sim \ln 7$ is specific for the hyper-cubic lattice
in $d=4$ and for charged particles (closed trajectories).

In other words, the condensation occurs again at the point
where the monopole mass vanishes, $m^2_{phys}=0$.
Moreover, the mass can be fine tuned to zero
by choosing the corresponding value of the electric charge
(\ref{condition}).

\subsection{No-go arguments for the $SU(2)$ case}

Imagine now that we would try to extract lessons from the Abelian cases and
suggest a mechanism for having a light scalar particle (monopole) 
in the non-Abelian case.
We would fail completely and come actually to a kind of a no-go theorem
for the non-Abelian case. 

{\it Mild fields.} It is reasonable to consider separately 
 mild  and hard gluonic fields. 
By mild fields we understand an analog of 
the Polyakov- 't Hooft monopole while by hard fields we understand
an analog of the monopoles of the compact (lattice) $U(1)$. 

There is no Higgs field in our case. 
Monopoles can still be defined following  Ref. \cite{thooft}. 
Namely,  pick up 
any scalar
field in the triplet representation, $H^a$. Then
vanishing of this field,
\beq
H^a~=~0~,~(a=1,2,3)
\eeq
represent three conditions specifying a world line
in $d=4$ space. The central point is that there
exists such a definition of the monopole charge that this trajectory
is nothing else but the monopole trajectory  \cite{thooft}. 

The problem is to identify, in absence of matter,
 a suitable $H^a$ field.
Indeed, locally the
field $H^a$ is to reduce to a product of non-Abelian field 
strength tensors $G^a_{\mu\nu}$. The simplest possibility is to choose
the Higgs field proportional to a particular component of the field
strength tensor, say,
$$H^a~=~G^a_{12}~.$$
 This choice would violate Lorentz invariance, however. If we try 
a Lorentz scalar then the simplest combination vanishes identically:
$$\epsilon^{abc}G^b_{\mu\nu}G^c_{\mu\nu}~\equiv~0~.$$
Increasing the number of the $G's$ in the product makes, first of
all, the construction unrealistically complicated. Moreover,
a more detailed consideration shows that it does not help at all
\cite{kovner}.

Thus, there is no way to construct a light monopole in terms of mild,
i.e. nonsingular fields.

{\it Hard fields}. Try now to construct a  fine tuned monopole, 
in analogy with the
lattice $U(1)$ theory. In the $U(1)$ case the price paid for
the absence of the Higgs field is    
singularities of the gauge field.
In particular, the magnetic current is
defined in terms of violations of the Bianchi identities,
\beq\label{definition}
j^{mon}_{\mu}~\sim~\epsilon_{\mu\nu\rho\sigma}\partial_{\nu}F_{\rho\sigma}~.
\eeq
Violation of the Bianchi identities assumes the fields to be
singular. However, all the expressions are regularized and 
well defined on the 
lattice \cite{degrand}.

In the $SU(2)$ case  definition (\ref{definition}) can be generalized to
a covariant form in an obvious way. Violation of the
Bianchi identities means that  
the non-Abelian fields are singular in the continuum
limit $a\to 0$. Singular fields imply a singular action.
Singular action means that the corresponding fluctuation is removed
from the physical spectrum unless the radiative mass is fine tuned to 
the entropy factor. So far, everything is similar to
 the $U(1)$ pure gauge theory. At this point, 
the analogy breaks down, however. A non-Abelian theory  is an
interacting
theory even in the absence of matter. As a result the coupling is
running and cannot be used as a parameter to tune the
propagating mass (\ref{physical}).

Thus, we come to the conclusion that neither finite nor singular 
non-Abelian fields can ensure appearance of a light scalar particle
(monopole).

\subsection{Lattice monopoles}

The no-go arguments above have been  realized
rather recently. Moreover,
the phenomenological studies of  
the lattice monopoles flourish, for a review see, e.g.,
\cite{suganuma}. Now, however, we can look backwards and ask,
why the no-go arguments did not work. This question
can be answered quantitatively, by measuring directly the
entropy and {\it non-Abelian} action of the monopoles
(we always mean monopoles of the maximal Abelian projection, see,
e.g., \cite{suganuma}). For the monopole action
the result of measurements is \cite{anatomy}:
\beq\label{S}
S_{mon}~\approx~\ln 7\cdot {L\over a}~,
\eeq
and for the entropy one gets \cite{boyko1}:
\beq\label{N}
N_L~\approx~(L/a)^7~~.
\eeq

We do not discuss here error bars in Eqs (\ref{S}),
(\ref{N}): details can be found in the original papers.
However, the answer to our question is quite clear,
compare (\ref{S}), (\ref{N}) and (\ref{physical}). Namely,
the monopoles exist because of a self-tuning. Both the action
and entropy are ultraviolet divergent but cancel each other
so that the propagating monopole mass is controlled by
$\Lambda_{QCD}^{-1}$ \cite{vz}. Note that the action is tuned to
a geometrical factor. On the other hand, 
the action of a pure Abelian monopole (embedded
into
the $SU(2)$) would be proportional to $g^{-2}(a)$
and could not be tuned to $\ln 7$ for varying $a$.
The mechanism of the self-tuning
in terms of the original Yang-Mills fields is not known.
However, it follows from the arguments above that this self-tuning
was the only ``chance'' for the monopoles to exist.

The self-tuning of the monopoles can be tested further, through
studies of properties of the monopole clusters.
In particular, one can predict the spectrum of finite clusters, $P(L)$ 
and their radii as
functions of their length \cite{maxim}:
\beq\label{nl}
P(L)~\sim~L^{-3}~, ~~R(L)~\sim~\sqrt{L\cdot a}~.
\eeq
The prediction is known to be true on the lattice \cite{hart,boyko}.

The main idea behind (\ref{nl})
is to use relations common to theory of percolating
systems. For further applications of the 
percolation theory to 
monopoles see \cite{boyko,ishiguro}.
Physicswise, what is most intriguing 
about the success of the
percolation
theory is that it treats monopoles as point like at the scale
of $a$. Thus, the monopoles emerge as a new probe
of short distances.

\section{Self-tuning of P-vortices}

\subsection{No-go arguments, second round}

Although the self-tuning of the monopoles allows to
describe a lot of data on the monopole clusters
\cite{maxim,boyko,ishiguro}, it brings us to 
a new question which looks even more devastating than the
original puzzles. Namely, if relations (\ref{S}) and (\ref{N})
is the ultimate truth about the lattice monopoles, then according
to (\ref{physical}) we have a  light scalar particle.
But 
because of the asymptotic freedom we may
have only three original gluons at short distances.

To scrutinize the argument we should translate the
problem into the language of the observables, that is
monopole trajectories.
 Note, first of all,  that point-like
particles in field theory are counted through ultraviolet
divergences.
Usually, in the continuum theory one deals with  logarithmic divergences.
In particular, the fact that there are only gluons at short distances
can be checked by studying the running of the coupling. 

However, the logarithmic running is a relatively small effect,
difficult to resolve. Luckily enough, on the 
lattice one can measure also  power-like divergences since the
ultraviolet
regularization is known explicitly. In particular, one can count
 particles by measuring the average
plaquette action. The standard prediction is:
\beq\label{plaquette}
\langle (G_{\mu\nu}^a)^2\rangle~\approx ~{(N_c^2-1)\over a^4}
\eeq 
Indeed, in the limit $a\to 0$ the plaquette action is to
 count zero-point fluctuations of gluons. 
Prediction (\ref{plaquette}) is easy to check.
Moreover, the relation (\ref{plaquette})
has perturbative corrections calculable, again, in terms of gluons
alone. In fact, many perturbative terms are explicitly known \cite{rakow}.

The first non-perturbative contribution to the plaquette
action  which is allowed by the continuum theory is that
of  the ultraviolet renormalon, for a recent
review and further references
see \cite{vz1}. This observation severely constrains the monopole contribution 
to the plaquette action:
\beq\label{plaquette1}
\langle (G_{\mu\nu}^a)^2\rangle_{mon}~~\sim~~\Lambda^2_{QCD}\cdot a^{-2}~.
\eeq

To appreciate implications of (\ref{plaquette1}) 
for the monopole trajectories we need to recollect
some elements of the geometrical picture. Monopoles occupy
centers of elementary cubes on the lattice. 
The cubes add up to trajectories.
The action, associated with the
monopole is measured as excess of action density
on the plaquettes forming the cubes times the volume of the cube, $a^3$.
In particular, (\ref{S}) means that the excess of the action
density on the plaquettes belonging to the monopoles  is
comparable to action density of the zero-point fluctuations 
(\ref{plaquette}).

Now, for the purpose of orientation 
 consider the standard model of percolation,
see, e.g., \cite{grimmelt}. One 
introduces probability $p$ for a cube to belong to a monopole
trajectory. At some value $p=p_{cr}$ the trajectories begin to
percolate,
i.e. there appears  a single cluster which spreads over the whole lattice.
Such a phenomenon is observed for the lattice monopoles indeed.
However, in this model, the total number of monopole cubes would be a finite
fraction, $p$ of all the cubes on the lattice.  As a result, the excess
of the action due to the monopoles, averaged over the lattice 
 would be comparable to 
(\ref{plaquette}). and 
the monopoles would interfere with the counting 
degrees
of freedom. We uncover  again the danger
that observations (\ref{S}), (\ref{N}) signify presence of new particles
at short distances, in contradiction with the asymptotic
freedom.

Thus, ordinary percolation is not allowed for the monopoles.
And what is allowed? Introduce probability
$\theta(cube)$
for an arbitrary cube on the lattice to contain a monopole. 
Then (\ref{plaquette1}) implies:
\beq\label{cube}
\theta~(cube)~\sim ~\big(a\cdot \Lambda_{QCD}\big)^2~~.
\eeq
That is, the probability for a cube to contain a monopole
with ultraviolet divergent action
should itself depend on $a$. The prediction might look even absurd at
first
sight but the drama is that
the data at presently available lattices \cite{boyko,muller} do comply with 
(\ref{cube}).

\subsection{Branes}

Although observation of (\ref{cube}) on the lattice settles our problem
with asymptotic freedom, the overall picture  might look
still very puzzling. Namely, the lattice monopoles share the properties
(\ref{S}), (\ref{N}) with a free particle in
$d=4$ space. On the other hand, the probability (\ref{cube}) depends
on the lattice spacing $a$, in striking distinction 
from the free particle in $d=4$.
The geometrical meaning of (\ref{cube}) is actually
transparent: monopoles live on a $d=2$ subspace of the whole $d=4$ space.
Indeed, the factor $(a\cdot \Lambda_{QCD})^2$ gives then the weight of
the cubes which belong to the surface provided that the area of the
surface is in physical units.

The conclusion might sound bizarre but actually phenomenologically this
picture, at least partially, has been known.
Namely, the surfaces are nothing else but the P-vortices, for review
see \cite{greensite}. Their total area scales in
physical units \cite{langfeld,greensite}. According to the latest data 
\cite{kovalenko}:
\beq\label{scaling}
A_{vort}~\approx~24~(fm)^{-2}\cdot V_4~,
\eeq
where $V_4$ is the lattice volume.  
Moreover it is known from the numerical simulations
that monopoles are strongly correlated with the P-vortices.
This was observed first for a single value of the lattice spacing
$a$ \cite{giedt} and
later confirmed for all the values of $a$ available \cite{syritsyn}.

Although scaling (\ref{scaling})  and conspiracy of the P-vortices
and monopoles were known since some time, no self-tuning was discussed 
until very recently \cite{kovalenko}. 
The missing point was the measurement of the vortex
action
as function of the lattice spacing $a$. The (non-Abelian) action
turned to be ultraviolet divergent \cite{kovalenko}:
\beq\label{area}
S_{vort}~\approx~0.54{A\over a^2}~.
\eeq

Observation (\ref{area}) means that the self-tuning is 
generalized from the monopoles to P-vortices \cite{kovalenko}.
Moreover, conspiracy of the monopoles and P-vortices 
 is elevated from a simple
phenomenological observation to a consequence of
the asymptotic freedom. 

\subsection{Interplay of the monopoles and vortices}

Let us emphasize that it would be an oversimplification to
say that the vortices are primary and monopoles are secondary.
The interplay between the monopoles and vortices is subtler.
Namely, at short distances monopoles `know' about $d=4$.
Indeed, the spectrum of finite clusters (\ref{nl}) is sensitive to the
number of dimensions:
\beq
N(L)~\sim ~L^{-1-d/2}~~,
\eeq
and the data indicate $d=4$, \cite{maxim,boyko}. 
Also, it is the monopole action which is tuned to the entropy 
of a point-like particle in $d=4$, compare (\ref{S}) and (\ref{N}).
Note that both the action and entropy are decided at short distances.
There is no such match for  the vortices if one would assume that
the vortices are structureless \cite{kovalenko}.

On the other hand, in the infrared the monopoles are associated with
a $d=2$ subspace of the full $d=4$ space. Thus, it is condensation
of the branes that is observed, not of monopoles.

\section{Conclusions}

The lattice data indicate that there exist branes in the
vacuum state of the lattice $SU(2)$ gluodynamics
which possess remarkable properties:

(a) total area scales in physical units;

(b) non-Abelian action is ultraviolet divergent;

(c) branes are populated with scalar particles, monopoles;

(d) monopoles are particles living on a $d=2$ subspace;

(e) the branes percolate and condense.

The branes are natural candidates for dual variables.

An important caveat is that in all the cases the evidence is 
entirely numerical and is, therefore,  true for a limited range of the
lattice spacings  available now. 
The only theoretical support for the  picture observed
we could find is
that if existence of a dual description were granted then
the basic features of the branes  listed above would be uniquely
fixed. 

Appearance of the branes might be related to a very special status
of dimension two condensates in gauge theories
, see, in particular, \cite{kondo}. 
Naively, the first gauge-invariant condensate,
$<(G_{\mu\nu}^2>$ has dimension four. The only Lorentz invariant
operator of dimension two, 
$(A_{\mu}^a)^2$ is seemingly gauge dependent. However,
one can argue that the minimal value of $<(A_{\mu}^a)^2>$, 
over the gauge orbit is gauge invariant. This minimization condition
implies non-locality. In case of pure electromagnetic field
\beq
<A^2_{\mu}>_{min}~=~(const)\int {d^4xd^4x^{'}\over
(x-x^{'})^2}F_{\mu\nu}(x)F_{\mu\nu}(x^{'})~~.
\eeq
In the non-Abelian case there is no such closed relation but an
explicit calculation in the Hamiltonian formalizm
\cite{vanbaal} demonstrates sensitivity of
the dimension-two condensate to the Gribov horizon.

Moreover, one can readily argue that the ultraviolet divergent part
 of $<(A^a_{\mu})^2>_{min}$ cannot be relevant to physical observables.
Thus, we are left with the non-perturbative part of
 $<(A^a_{\mu})^2>_{min}$
which is of order $\Lambda_{QCD}^2$. As we argued in \cite{vz} the monopoles
realize
\beq
<|\phi_M|^2>~\sim~\Lambda_{QCD}^2~~,
\eeq
where $\phi_M$ is magnetically charged field.
Properties of the branes seem to reflect the balance of locality and
non-locality
inherent to the gauge invariant operator of dimension 2 in gauge
theories.

\section*{Acknowledgments}

This note grew out of discussions of the results of the lattice
measurements presented in 
\cite{anatomy,kovalenko,syritsyn,boyko1,boyko}. 
I am thankful to V.G. Bornyakov,
P.Yu. Boyko, M.N. Chernodub, F.V. Gubarev, A.V. Kovalenko, M.I. Polikarpov, 
T. Suzuki and S.N. Syritsyn for collaboration and thorough discussions.

I am thankful to A. Bartl, M. Faber,
A. Di Giacomo, J. Greensite, J. Kubo, P. van Nieuwenhuizen, 
A.A. Slavnov, L. Stodolsky, H. Terao and E.T. Tomboulis 
for interesting discussions. 

The paper was completed while the author was visiting the Yukawa
Institute for Theoretical Physics of the Kyoto University.
I am thankful to Prof. T. Kugo and other members of the group for
the hospitality. 

The material of this note was presented as a talk
at the Conference ``Confinement 2003'' at RIKEN (Japan), July 21-24,
(2003)  and as
a lecture at the Summer Institute, Fuji-Yoshida (Japan), August 12-19,
(2003). I am thankful to the organizers, Prof.
H. Suganuma and Prof. J. Kubo, for the invitations.

\end{document}